\documentclass[preprint,pra]{revtex4}%
\usepackage{amsfonts}
\usepackage{amsmath}
\usepackage{amssymb}
\usepackage{graphicx}%
\providecommand{\U}[1]{\protect\rule{.1in}{.1in}}

\begin{document}
\preprint{HEP/123-qed}
\title[Electric dipole moment searches: \ II]{Electric dipole moment searches: Effect of linear electric field frequency
shifts induced in confined gases, II}
\author{R. Golub}
\affiliation{Physics Dept, NCSU}
\author{C.M. Swank}
\affiliation{Physics Dept}
\affiliation{North Carolina State Universty,}
\affiliation{Riddick Hall}
\affiliation{Raleigh, NC 27610}
\author{S.K. Lamoreaux}
\affiliation{Yale University}
\affiliation{Department of Physics}
\affiliation{PO Box 208120}
\affiliation{New Haven, Conn. 06520}
\pacs{PACS number}

\begin{abstract}
The next generation of particle edm searches will be at such a high
sensitivity that it will be possible for the results to be contaminated by a
systematic error resulting from the interaction of the motional ($E\times
v/c$) magnetic field with stray field gradients. In this paper we extend
previous work to present an analytic form for the frequency shift in the case
of a rectangular storage vessel and discuss the implications of the result for
the neutron edm experiment which will be installed at the SNS (Spallation
Neutron Source) by the LANL\ collaboration

\end{abstract}
\volumeyear{year}
\volumenumber{number}
\issuenumber{number}
\eid{identifier}
\date[Date text]{date}
\received[Received text]{date}

\revised[Revised text]{date}

\accepted[Accepted text]{date}

\published[Published text]{date}

\startpage{101}
\endpage{ }
\maketitle
\tableofcontents

\section{Introduction}

Searches for particle electric dipole moments (edm) are considered to be one
of the most promising places to search for physics beyond the standard model.
Current experiments have reached the sensitivity where they have to take into
account a systematic effect due to the influence on the particle's magnetic
dipole moment of a combination of the motional magnetic field, $\left(
\overrightarrow{B}_{m}=\overrightarrow{E}\times\overrightarrow{v}/c\right)  $
due to the motion of the particle in the applied static electric field, and
gradients in the ambient magnetic field. For slow motion (adiabatic limit) the
effect can be described as a geometric phase effect \cite{COM} whereas a
description in terms of the Bloch-Siegert shift is valid for fast motions as
well (\cite{JMP}). A treatment valid also for intermediate motions for
particles moving in a cylindrical container has been given in \cite{SKL},
\cite{BAR}. In this note we discuss some additional symmetries of the effect
and present an analytic solution for the case of particles moving in a
rectangular vessel.

\section{Symmetry of the effect}

It is easy to see \cite{JMP}, that the effect depends only on motion in a
plane perpendicular to the direction of $\overrightarrow{E}$ (parallel motion
produces no $\overrightarrow{B}_{m}$). Equations (23 and 36) in \cite{SKL}
show that in general the frequency shift linear in $\overrightarrow{E}$ which
leads to the systematic error is given in terms of the spectrum of the
velocity autocorrelation function by
\begin{equation}
\Delta\Omega_{E}=-\frac{\gamma^{2}E}{2c}\left[  \frac{\partial B_{y}}{\partial
y}S_{y}\left(  \omega_{o}\right)  +\frac{\partial B_{x}}{\partial x}%
S_{x}\left(  \omega_{o}\right)  \right]  \label{one}%
\end{equation}
with%
\begin{align}
S_{i}\left(  \omega_{o}\right)   &  =\int_{0}^{\infty}\cos\omega_{o}\tau
R_{i}\left(  \tau\right)  d\tau\\
R_{i}\left(  \tau\right)   &  =2\int_{0}^{\tau}\psi_{i}\left(  x\right)  dx\\
\psi_{i}\left(  x\right)   &  =\left\langle v_{i}\left(  t\right)
v_{i}\left(  t-x\right)  \right\rangle \\
S_{i}\left(  \omega_{o}\right)   &  =2\int_{-\infty}^{\infty}\frac{\Psi
_{i}\left(  \omega\right)  }{\left(  \omega_{0}^{2}-\omega^{2}\right)
}d\omega
\end{align}
where $\omega_{o}=\gamma B_{o}$ is the Larmor frequency, $\gamma$ is the
gyromagnetic ratio, $B_{o}$ is the homogeneous magnetic field taken as
parallel to $\overrightarrow{E}$ in the $z$ direction and $\Psi_{i}\left(
\omega\right)  $ is defined by%
\begin{equation}
\psi_{i}\left(  x\right)  =\int_{-\infty}^{\infty}\cos\omega x\Psi_{i}\left(
\omega\right)  d\omega
\end{equation}
. If the field is cylindrically symmetric%
\begin{align}
\frac{\partial B_{y}}{\partial y}  &  =\frac{\partial B_{x}}{\partial
x}=-\frac{1}{2}\frac{\partial B_{z}}{\partial z}\\
\Delta\Omega_{E}  &  =\frac{\gamma^{2}E}{2c}\frac{1}{2}\frac{\partial B_{z}%
}{\partial z}\left(  S_{y}\left(  \omega_{o}\right)  +S_{x}\left(  \omega
_{o}\right)  \right)  \label{two}%
\end{align}
and the result depends only on $\frac{\partial B_{z}}{\partial z}$ for all
frequencies and geometries of the orbits. In the case that the trajectories
have cylindrical symmetry\quad%
\begin{align}
S_{y}\left(  \omega_{o}\right)   &  =S_{x}\left(  \omega_{o}\right)  =S\left(
\omega_{o}\right) \label{3}\\
\Delta\Omega_{E}  &  =\frac{\gamma^{2}E}{2c}\frac{\partial B_{z}}{\partial
z}S\left(  \omega_{o}\right)
\end{align}

This symmetry will hold in the high frequency limit ($\omega_{o}>>\omega_{r}$)
which is determined by the short time behavior of the correlation function,
and the result that the shift in this case depends only on $\frac{\partial
B_{z}}{\partial z}$ has been obtained in \cite{JMP}, section IV B and by a
different method in \cite{BAR} section IV A. The advantage of the present
treatment is obvious.

An interesting case that can arise in practice is the symmetry%
\begin{align}
\frac{\partial B_{z}}{\partial z}  &  \approx0,\text{ \ }\frac{\partial B_{x}%
}{\partial x}\approx-\frac{\partial B_{y}}{\partial y}>>0\\
\Delta\Omega_{E}  &  =\frac{\partial B_{x}}{\partial x}\left(  S_{x}\left(
\omega_{o}\right)  -S_{y}\left(  \omega_{o}\right)  \right)
\end{align}
In the high frequency limit this difference in the spectra will approach zero.

\section{Motion in a rectangular box}

\subsection{Single velocity}

In a rectangular box with specular reflecting walls the orbits are straight
lines reflecting \ at equal angles to the normal when they encounter a wall.
With perpendicular walls parallel to the $x$ and $y$ axes the motion in each
dimension will be independent of the other dimension and the angle with the
normal will be preserved independently for the walls along $x$ and $y$. At a
wall collision the magnitude of velocity and the component parallel to the
wall are unchanged and the perpendicular component of velocity changes sign.
Thus the correlation function for a given velcocity component $\left(
i=x,y\right)  $ starts at $v_{i}^{2}$ and after $l$ collisions is%
\begin{equation}
\psi_{i}\left(  \tau\right)  \equiv\left\langle v_{i}\left(  \tau\right)
v_{i}\left(  0\right)  \right\rangle =v_{i}^{2}\left(  -1\right)  ^{l}\quad
for\quad lT_{i}<\tau<\left(  l+1\right)  T_{i}%
\end{equation}
with the time between collisions $T_{i}=L_{i}/v_{i}$, being constant for each
orbit and $L_{i}$ the length of the box in direction $i.$

Thus the velocity correlation function for each direction is a square wave
with the switching points being equally spaced but whose exact timing depends
on the distance of the starting point of a given orbit from the first wall
collision. Averaging over these starting points proceeds as in equations 14-17
of \cite{BAR}. In fact the orbits are exactly those characterized by
$\alpha=\pi/2$ in that reference (see fig.2 \cite{BAR}) and the results of
that paper can be applied to the present case by substituting
\begin{align}
\alpha &  =\pi/2\\
R  &  =L_{i}/2
\end{align}
Thus equ. 42 \cite{BAR} becomes%
\begin{equation}
S_{i}\left(  \omega_{o}\right)  =\frac{L_{i}^{2}}{4}\sum_{m=-\infty}^{\infty
}\frac{1}{\left(  \pi\left(  m+1/2\right)  \right)  ^{2}}\left[  \frac{\left(
\omega_{o}^{\prime\prime2}-\left(  \pi\left(  m+1/2\right)  \right)
^{2}\right)  }{\left(  \left(  \omega_{o}^{\prime\prime2}-\left(  \pi\left(
m+1/2\right)  \right)  ^{2}\right)  ^{2}+\omega_{o}^{\prime\prime2}r_{o}%
^{2}\right)  }\right]
\end{equation}
with $\omega_{o}^{\prime\prime}=\omega_{o}L_{i}/2v_{i}$ and $r_{o}%
=L_{i}/2\lambda$, $\lambda$ being the mean free path between gas collisions.
Introducing $\omega_{o}^{\prime}=\omega_{o}L_{i}/v_{i}$ we have%
\begin{equation}
S_{i}\left(  \omega_{o},l_{o}\right)  =L_{i}^{2}\sum_{m=-\infty}^{\infty}%
\frac{1}{\left(  \pi\left(  m+1/2\right)  \right)  ^{2}}\left[  \frac{\left(
\omega_{o}^{\prime2}-\left(  \pi\left(  2m+1\right)  \right)  ^{2}\right)
}{\left(  \left(  \omega_{o}^{\prime2}-\left(  \pi\left(  2m+1\right)
\right)  ^{2}\right)  ^{2}+\omega_{o}^{\prime2}l_{o}^{2}\right)  }\right]
\label{4}%
\end{equation}
with $l_{o}=\frac{L_{i}}{\lambda}$. This is plotted in figure 1), which is to
be compared with figure 3) of \cite{BAR} for the case of a cylinder.

Equation (\ref{4}) and figure 1) apply to motion in one dimension. For the
full 2 dimensional problem it is necessary to add together suitably normalized
forms of the function for each dimension.

The value of $S_{i}\left(  \omega_{o}=0\right)  =.083L_{i}^{2}$ agrees with
that predicted by the diffusion theory for the single dimension contribution
to a rectangular box (eqation 82, \cite{SKL}),
\begin{equation}
\frac{8}{\pi^{4}}\sum_{m=1,3,5}\frac{1}{m^{4}}=.083
\end{equation}
Equation \ref{4} shows that $S_{i}\left(  \omega_{o}\right)  $ becomes
independent of $L_{i}$ as the frequency increases and the larger the damping
(larger $l_{o}$), the higher the frequency where this occurs.

\subsection{Frequency shift averaged over Maxwell velocity distribution, the
case of co-magnetometers.}

Both the neutron edm experiment carried out by Baker \emph{et al,
}\cite{JMPedm} and that being developed by the Los Alamos collaboration
\cite{LANL}, make use of co-magnetometers, that is a gas of atoms occupying
the same region as the ultra-cold neutrons and satisfying the Maxwell-Bolzmann
velocity distribution. In this case we write the velocity as
\begin{equation}
v_{i}=y\beta(T)
\end{equation}
with $\beta\left(  T\right)  =\sqrt{\frac{2kT}{m}}$ the most probable velocity
in a volume. Then our one dimensional velocity $v_{i}$ has the probability
distribution%
\begin{equation}
P\left(  y\right)  dy=\frac{2}{\sqrt{\pi}}e^{-y^{2}}dy
\end{equation}
The spectral function of the frequency shift can then be rewritten%
\begin{equation}
S_{i}\left(  \omega_{o},y,l_{o}^{\ast}\right)  =L_{i}^{2}\sum_{m=-\infty
}^{\infty}\frac{1}{\left(  \pi\left(  m+1/2\right)  \right)  ^{2}}\left[
\frac{\left(  \omega_{o}^{\ast2}-\left(  \pi\left(  2m+1\right)  \right)
^{2}y^{2}\right)  y^{2}}{\left(  \left(  \omega_{o}^{\ast2}-\left(  \pi\left(
2m+1\right)  \right)  ^{2}y^{2}\right)  ^{2}+\omega_{o}^{\ast2}\left(
l_{o}^{\ast}\right)  ^{2}\right)  }\right]
\end{equation}
where $\omega_{o}^{\ast}=\omega_{o}L_{i}/\beta\left(  T\right)  ,$ and
$l_{o}^{\ast}=L_{i}/\left(  \beta\left(  T\right)  \tau_{c}\left(  T\right)
\right)  $ and we have specialized to the case where the gas collision time,
$\tau_{c}$ is independent of velocity. This is valid for the common case where
the scattering cross section satisfies $\sigma\sim1/v$ and holds in particular
for the case of $He^{3}$ diffusing in superfluid $He^{4}$ which is the
co-magnetometer in the LANL experiment \cite{LANL}, \cite{PHYSrep}.

\subsubsection{$He^{3}$ colliding with phonons in superfluid $He^{4}$, the
co-magnetometer in the LANL search for a neutron electric dipole moment.}

Since the velocity of the $He^{3}$ is much less than the phonon velocity, the
collision rate of the phonons with the $He^{3}$ will be independent of
velocity and the mean free path satisfies $\lambda=v\tau_{c}$. We obtain
$\tau_{c}$ from
\begin{equation}
\tau_{c}\left(  T\right)  =3D\left(  T\right)  /\left\langle v^{2}%
\right\rangle _{T}%
\end{equation}
with $\left\langle v^{2}\right\rangle _{T}$ the mean square velocity in a
volume of gas and $D\left(  T\right)  $ has been measured at temperatures of
interest \cite{D3}%
\begin{equation}
D\left(  T\right)  =\frac{1.6}{T^{7}}cm^{2}/\sec
\end{equation}
We now average the frequency shift over the Maxwell- Boltzman distribution for
velocity in one dimension%
\begin{equation}
\Psi_{x}\left(  \omega_{o}^{\ast},T\right)  =-\frac{2}{L_{i}^{2}\sqrt{\pi}%
}\int e^{-y^{2}}S_{i}\left(  \omega_{o}^{\ast},y,l_{o}^{\ast}\right)  dy
\label{5}%
\end{equation}
The results are plotted in figure 2):

The same result is shown as a function of temperature for fixed (normalized)
frequency in fgure 3. (Note the frequency normalization is temperature dependent.

\subsection{A rectangular box with the two sides significantly different,}

Equations \ref{4} and \ref{5} refer to \ a single dimension. If the second
dimension of the box has a length $L_{y}$, then equation 4 can be written
$\left(  \varepsilon=L_{y}/L_{x}\right)  ,$ keeping the same normalization for
$S_{y}$ and $\omega_{o}^{\ast}$:%
\begin{equation}
\frac{S_{y}\left(  \omega_{o},y,l_{o}^{\ast}\right)  }{L_{x}^{2}}%
=\varepsilon^{2}\sum_{m=-\infty}^{\infty}\frac{1}{\left(  \pi\left(
m+1/2\right)  \right)  ^{2}}\left[  \frac{\left(  \omega_{o}^{\ast2}-\left(
\pi\left(  2m+1\right)  \right)  ^{2}y^{2}\right)  y^{2}}{\left(  \left(
\varepsilon^{2}\omega_{o}^{\ast2}-\left(  \pi\left(  2m+1\right)  \right)
^{2}y^{2}\right)  ^{2}+\omega_{o}^{\ast2}\left(  l_{o}^{\ast}\right)
^{2}\varepsilon^{2}\right)  }\right]
\end{equation}

Averaging this as in equation \ref{5} we obtain, $\left(  \Psi_{y}\left(
\omega_{o}^{\ast},T,l_{o}^{\ast}\right)  =\Psi_{x}\left(  \varepsilon
\omega_{o}^{\ast},T,l_{o}^{\ast}\varepsilon\right)  \right)  $
\begin{equation}
\Psi\left(  \omega_{o}^{\ast},T\right)  =\Psi_{x}\left(  \omega_{o}^{\ast
},T\right)  +\varepsilon^{2}\Psi_{y}\left(  \omega_{o}^{\ast},T\right)
\end{equation}

Figure 4) shows the two dimensional result for $\varepsilon=0.2,$ normalized
to $L_{x}^{2}$ as a function of frequency normalized to $L_{x}$ for various
temperatures while figure 5) shows the normalized shift vs temperature for
various normalized frequencies.

In figure 6) we show the contribution to the frequency shift for the two
directions independently, compared to the results of numerical simulations, as
well as the total frequency shift given by their sum for T=0.4K. The
contribution of the short side ($y$) has been normalized to the long side
($x$) for the case $\varepsilon=L_{y}/L_{x}=0.2.$ We are assuming, for this
discussion that the two components of the gradients are equal%
\begin{equation}
\frac{\partial B_{o}}{\partial x}=\frac{\partial B_{o}}{\partial y}=\frac
{1}{2}\frac{\partial B_{o}}{\partial z}%
\end{equation}
Otherwise each curve in fig. 6 will have to be multiplied by the appropriate
gradient and the total effect will be altered.

We see that around the zero crossing the contributions of the two dimensions
contribute with opposite sign so the symmetry (equation \ref{3}) does not hold
and if we wish to operate near the zero crossing the result will depend on
$\frac{\partial B_{x}}{\partial x}$ and $\frac{\partial B_{y}}{\partial y}$
separately and not on $\frac{\partial B_{z}}{\partial z}$ except in the case
of cylindrical symmetry $\frac{\partial B_{o}}{\partial x}=\frac{\partial
B_{o}}{\partial y}=\frac{1}{2}\frac{\partial B_{o}}{\partial z}.$

Finally fig.7 shows the comparison of theory and numerical simulation for T=300mK.

\section{Discussion}

The calculations of the velocity correlation function (vcf) presented here and
in \cite{BAR}, start by following a single trajectory with the collisions only
damping the amplitude of the vcf and not changing the velocity components,
$v_{x},v_{y}$ as the particle moves along this trajectory. The simulations, on
the other hand follow a particle as it is deflected to another (randomly
chosen) trajectory by the collisions. Essentially the theory follows the
particles that haven't collided while the simulations follow those that have.

Each collision, while conserving the kinetic energy (magnitude of velocity),
will result in a change of direction of the motion and hence of the $x$ and
\ $y$ components of velocity so the correlation function resulting from the
simulation of a single trajectory will be quite different from that considered
by the theoretical calculation. However, for each collision that takes a
particle from trajectory (1) to trajectory (2) there should be a collision
leading to the reverse transition according to detailed balance and when we
average over all trajectories the results are seen to be the same.

In addition while we have neglected the motion parallel to the E field we see
that collisions will alter the velocity component in this direction, and, as a
result of conservation of energy in the collisions, will thus alter the
velocity in the perpendicular plane. However this effect will be cancelled
when we average over all trajectories in the perpendicular plane as we have
done here and in \cite{BAR}. (We note that due to the heavy mass and slow
$He^{3}$ velocity, Baym and Ebner \cite{BE} conclude that the phonon
scattering on $He^{3}$ is predominantly elastic.)

We have presented the general solution for the frequency shift linear in E,
for the case of a rectangular box with specular reflecting walls. The effects
of non-specular wall reflections are expected to be small for the case of
heavy damping of interest with respect to the co-magnetometers. For UCN\ the
non-specular wall collisions are expected to be the major source of damping
\ but the effect is expected to be small. This will be discussed in a
subsequent work.

\bigskip

\section{\bigskip Figure Captions}

Fig. 1) Normalized frequency shift vs normalized frequency, $\omega^{\prime
}=\omega_{o}L_{x}/v,$ for the single dimension contribution of a single velocity.

Fig. 2) One dimensional contribution to the normalized velocity averaged
frequency shift vs. reduced frequency $\omega_{x}=\omega_{o}L_{x}/\beta(T)$,
for Temperatures T=0.1, 0.2 0.3 and 0.4K, using the temperature dependent mean
free path for $He^{3}$ in $He^{4}$

Fig. 3) One dimensional contribution to normalized velocity averaged frequency
shift with normalized frequency ($\omega^{\ast}=\omega L_{i}/\beta(T)$) as a parameter.

Fig. 4) Normalized frequency shift for a rectangular box with $\varepsilon
=L_{y}/L_{x}=0.2$ normalized to $L_{x}^{2}$ vs frequency normalized to
$L_{x}/\beta\left(  T\right)  $ for various temperatures.

Fig. 5) Frequency shift in a rectangular box with $\varepsilon=L_{y}%
/L_{x}-0.2$ normalized to $L_{x}^{2}$ vs temperature, $K$, with normalized
frequency as a parameter.

Fig. 6) Contribution of the long dimension, $x,$ (red),short dimension, $y,$
(blue) and combned result (violet) for the frequency shift normalized to long
dimension $\left(  x\right)  $,vs, frequency normalized to the longdimension,
$\omega1=\omega^{\ast}=\omega_{o}L_{x}/\beta(T)$ for a temperature of 0.4K and
$\varepsilon=L_{y}/L_{x}=0.2$ The results of the theory are shown along with
those derived by numerical simulations.

Fig. 7) Contribution of the long dimension, $x,$ (green),short dimension, $y,$
(red) and combned result (violet) for the frequency shift normalized to long
dimension $\left(  x\right)  $,vs, frequency normalized to the longdimension,
$\omega^{\ast}=\omega_{o}L_{x}/\beta(T)$ for a temperature of 0.3K and
$\varepsilon=L_{y}/L_{x}=0.2$ The results of the theory are shown along with
those derived by numerical simulations.

\end{document}